\documentclass[aps,preprint,prd,superscriptaddress,nofootinbib]{revtex4}

\usepackage{amsmath,amssymb}
\usepackage{graphicx}
\graphicspath{ {./figures/} }
\usepackage{aas_macros}
\usepackage{subcaption}
\usepackage{float}
\RequirePackage[colorlinks,citecolor=blue,urlcolor=blue]{hyperref}
\usepackage{soul}
\usepackage{doi}
\usepackage{hyperref}  % 允许生成超链接

\usepackage{placeins}
\usepackage{fontspec}
\usepackage{xeCJK}
\usepackage{booktabs}

\begin{document}

\title{Searching for gravitational wave echo with group equivariant neural posterior estimation}

\author{Jian-Wei Luo\footnote{\href{luojianwei24@mails.ucas.ac.cn}{luojianwei24@mails.ucas.ac.cn}}}
\affiliation{School of Fundamental Physics and Mathematical Sciences,Hangzhou Institute for Advanced Study, UCAS, Hangzhou 310024, China}
\affiliation{Institute of Theoretical Physics, Chinese Academy of Sciences, Beijing 100190, China}
\affiliation{University of Chinese Academy of Sciences, Beijing 100190, China}

\author{Yun-Song Piao\footnote{\href{yspiao@ucas.ac.cn}{yspiao@ucas.ac.cn}}}
\affiliation{School of Fundamental Physics and Mathematical Sciences,Hangzhou Institute for Advanced Study, UCAS, Hangzhou 310024, China}
\affiliation{Institute of Theoretical Physics, Chinese Academy of Sciences, Beijing 100190, China}
\affiliation{School of Physical Sciences, University of Chinese Academy of Sciences, Beijing 100049, China}
\affiliation{International Center for Theoretical Physics Asia-Pacific, Beijing/Hangzhou, China}

%\author{author$^{1}$\footnote{email@example.com}}

%\affiliation{$^1$ some affiliations}

\begin{abstract}

Current methods for the gravitational-wave (GW) echo searches
generally require substantial computational time and resources.
Recent advances in group equivariant neural posterior estimation
(GNPE) have demonstrated the potential to substantially accelerate
posterior inference. In this work, we adopt this framework and
train a GNPE model to perform echo searches,  significantly
improving computational efficiency while maintaining comparable
inference accuracy. We apply it to the O4a events and find no
significant evidence for the GW echo.
\end{abstract}

\maketitle

%\tableofcontents

\section{Introduction}

Since 2015, the LIGO-Virgo-KAGRA collaboration has detected
hundreds of gravitational wave (GW)
events~\cite{nitz20234,ligo2023gwtc,abac2026gwtc4_1,abac2026gwtc4_2,abac2026gwtc,abac2605gwtc,abac2026gwtc5},
opening a new window into the strong-gravity regime. These events
can be interpreted as mergers of binary black holes (BBHs) or
binary neutron stars (BNSs) \cite{mandel2022rates}.

%alternative mechanisms for generating gravitational waves cannot
%be ruled out.

In particular, exotic compact objects (ECOs) without
horizons\cite{visser2009small,cardoso2016gravitational,cardoso2017testing,holdom2017not,zhang2018can,maggio2017exotic,galvez2019echoes,maselli2019micro},
such as boson
stars\cite{liebling2023dynamical,brito2017stochastic,palenzuela2017gravitational},
fuzzballs\cite{mathur2005fuzzball,muguruza2026black},
gravastars\cite{mazur2004gravitational,visser2004stable}, and
others~\cite{cardoso2017observational,cardoso2019testing}, may
also account for the GW
events~\cite{mark2017recipe,cardoso2019gravitational,zimmerman2023rogue}.
Because ECOs are not perfect absorbers, they are expected to
reflect GW near the
horizon~\cite{cardoso2016gravitational,cardoso2016gravitational2},
leading to a series of postmerger ``echoes'' following the main
ringdown\cite{cardoso2017observational,siemonsen2024nonlinear}. Echoes may also arise in alternative theories of gravity; for example, massive gravity can give rise to a characteristic double-peak potential that may produce gravitational-wave echoes in objects with horizons\cite{dong2021gravitational}.
The GW echo, if they can be detected, can be probes of new physics
at the near-horizon regime, which motivated the searching for the
corresponding signals in GW
data~\cite{abedi2017echoes,wang2019echo,wang2020searching,lai2026testing,lai2026gw190521,abbott2021tests,abbott2021tests2,uchikata2023searching,bhattacharjee2026gravitational}.

Current searches show no significant evidence for the GW echo.
This may be due to the expected echo signals are extremely weak.
For instance, the amplitude of echo is constrained to 15\% of the
merger amplitude for GW150914~\cite{nielsen2019parameter} and
another study shows that necessary signal-to-noise ratio of the
ringdown signal is 20–60 to detect echo~\cite{longo2021loud}. It
might be expected that with upcoming advances in technology and
improvements in detector sensitivity, such echo may possibly be
detected.

Deep learning has demonstrated remarkable efficacy in various GW
data analysis tasks, including signal
detection~\cite{george2018deep,gabbard2018matching,wang2020gravitational,krastev2020real,lopez2021deep},
parameter
estimation~\cite{gabbard2022bayesian,dax2021group,dax2021real} and
signal
extraction~\cite{torres2016denoising,wei2020gravitational,shen2019denoising,wang2024waveformer,chatterjee2021extraction,xu2024gravitational}.
Notably, Ref.~\cite{dax2021group} presented group equivariance
into normalizing flows. They showed that group equivariant neural
posterior estimation (GNPE) achieves state-of-the-art accuracy but
with far less inference times.

%Furthermore, machine-learning-based parameter estimation can
%substantially reduce computational cost and inference time.
In the study, we trained a GNPE model for detecting GW echo based
on the \texttt{dingo}
framework~\cite{green2020gravitational,green2021complete,dax2021real,dax2023neural,wildberger2023adapting},
and applied it to search for the echo in the O4a events. The
results show a significant improvement in computational
efficiency. In particular, our model can perform an echo search
for a single event in less than 20 minutes on a single RTX 5090.
%whereas traditional Bayesian inference methods typically require
%several hours.
Consistent with the previous analysis in Ref.~\cite{abac2026gwtc},
no significant evidence for the echo was observed.

\section{Methods}

\subsection*{Dataset}
Although the physics near the event horizon has not yet been fully
understood, several practical GW echo templates have already been
developed~\cite{abedi2017echoes,wang2019echo,Wang:2018mlp,Li:2019kwa,maselli2017parameter,wang2018black,ashton2016commentsonechoesabyss,testa2018analytical,abedi2017echoes2,burgess2018effective}.
Here we adopt a constant-interval echo
template~\cite{abedi2017echoes}, which involves five free
parameters: $\Delta t_{echo}$, $t_{echo}$, $t_0$, $\gamma$ and
$A$. The echo template $h_{TE}$ is defined as
\begin{align}
h_{TE}(t,t_0)&=A \sum^{\infty}_{n=0}(-1)^{n+1}\gamma^nh_T(t+t_{merger}-t_{echo}-n\Delta t_{echo},t_0),\\
h_T(t,t_0)&=\frac{1}{2}h(t)\{1+tanh[\frac{1}{2}\omega_I(t)(t-t_{merger}-t_0)]\}.
\label{template}\end{align} where $\omega_I(t)$ is model
frequency~\cite{abbott2016tests} and $t_{merger}$ is time of
merger. Fig.~\ref{fig:CIE} presents an example of an IMRPhenomXPHM
waveform with echo template (\ref{template}). For visualization
purposes, the amplitude parameter $A$ is set to 0.5, which is
larger than expected value.

%1. $\Delta t_{echo}$ is the time-interval between successive echoes.%,varied within the 1$\sigma$ range.
%
%2. $t_{echo}$ is the arrival time of the first echo.
%
%3. $t_0$ determines which part of the GR merger template is truncated. To achieve this, a  smooth cut-off function is introduced:
%\begin{equation}
%\Theta(t,t_0)=\frac{1}{2}\{1+tanh[\frac{1}{2}\omega_I(t)(t-t_{merger}-t_0)]\},
%\end{equation}
%where $\omega_I(t)$ is model frequency as a function of time~\cite{abbott2016tests} and $t_{merger}$ is time of merger. %, characterized by the peak of the template. %Generally,  $t_0$ is assumed to be negative and varied within the range $t_0\in(-0.1,0)\overline{\Delta t}_{echo}$.
% With the cut-off function, the truncated template from the waveform $h(t)$ can be define as
%\begin{equation}
%h_T(t,t_0)=\Theta(t,t_0)h(t).
%\end{equation}
%
%4. $\gamma$ is the damping factor between successive echoes.
%
%5. $A$ is the overall amplitude of the echo template relative to the main event.
%
%With these five parameters, the echo template is defined as
%\begin{equation}
%h_{TE}(t,t_0)=A \sum^{\infty}_{n=0}(-1)^{n+1}\gamma^nh_T(t+t_{merger}-t_{echo}-n\Delta t_{echo},t_0).
%\end{equation}

\begin{center}
\refstepcounter{figure}
\includegraphics[width=0.8\textwidth]{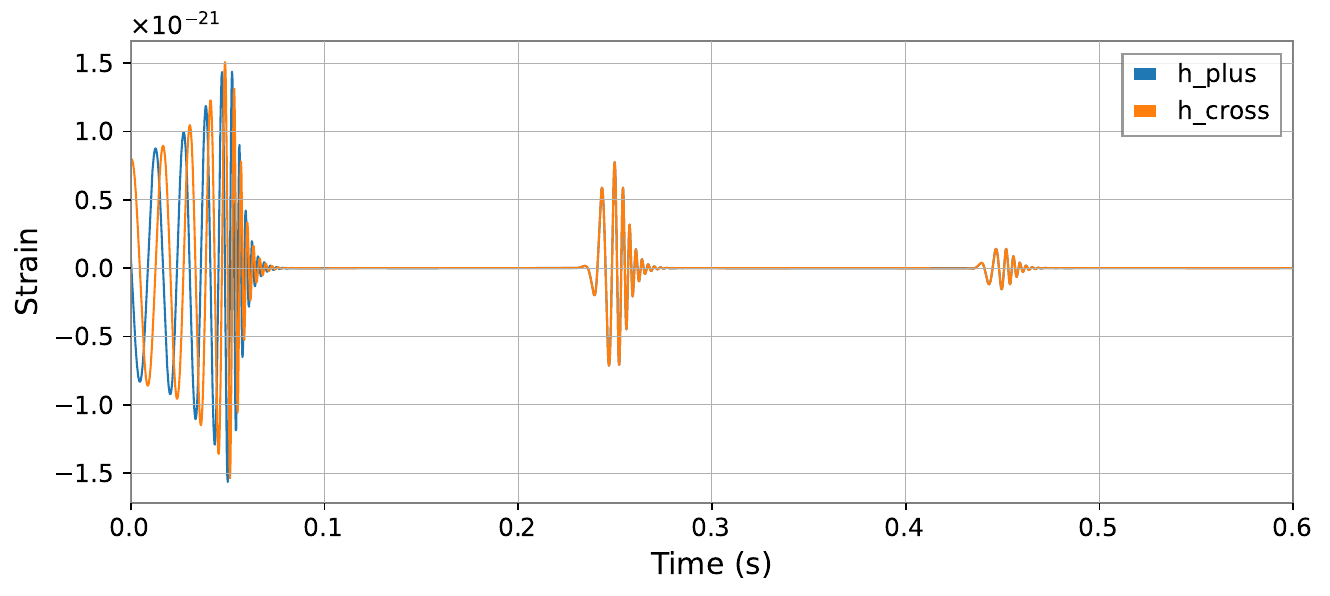}

\textbf{Figure \thefigure.}
Illustration of an IMRPhenomXPHM waveform with echo. %generated by CIE template.
\label{fig:CIE}
\end{center}

Using the waveform template IMRPhenomXPHM+Echo, we generated $5
\times 10^6 $ waveforms. We use the same prior over parameters,
with $m_1,m_2 \in [5,100] M_\odot$. The echo parameter priors are
listed in table~\ref{tab:priors}, consistent with those used in
Ref.~\cite{abac2026gwtc}, where $M^{maxL}$ in
Ref.~\cite{abac2026gwtc} is set to $50\,M_\odot$ so that
$t_M\approx 0.25 ms$.
%The noise amplitude spectral density (ASD) was derived from data of the O4a run available on the Gravitational Wave Open Science Center (GWOSC)~\cite{vallisneri2015ligo,ligo2021open,abbott2023open}. Together with the corresponding parameters of the generated waveforms, these constitute the training dataset.
\begin{table}[htbp]
\centering
\caption{Prior ranges of echo parameters.}
\label{tab:priors}
\begin{tabular}{cc||cc}
\hline
Parameter & Range & Parameter & Range \\
\hline
$\log_{10} A$ & $[-2,0]$ & $t_0$ & $[-100,10]t_M$ \\
$\log_{10} \gamma$ & $[-2,0]$ & $t_{echo}$ & $[10,10^3]t_M$ \\
$\Delta t_{echo}$ & $[10,10^3]t_M$ & & \\
\hline
\end{tabular}
\end{table}

During training, the noise signals during the O4a run are also part of the data. We downloaded all data from the H1 and L1 detectors during the O4a run from the Gravitational Wave Open Science Center (GWOSC) and extracted the PSD every 1024 s.

\subsection*{Group equivariant neural posterior estimation}

Neural posterior estimation
(NPE)~\cite{papamakarios2016fast,greenberg2019automatic} is a
simulation-based inference method to build Bayesian conditional
density estimation. Given a dataset of prior parameter samples
$\theta^{(i)} \sim p(\theta)$ and corresponding model simulations
$x^{(i)} \sim p(x|\theta^{(i)})$, it trains a neural posterior
estimation $q(\theta|x)$ to estimate $p(\theta|x)$ by minimizing
the loss function
\begin{align}
\mathcal{L}_{NPE} = \mathbb{E}_{p({\theta})}\mathbb{E}_{p({x|\theta})}[-log(q(\theta|x))]
\end{align}
over the dataset of $(\theta^{(i)}, x^{(i)})$ pairs. Normalizing
flows~\cite{rezende2015variational,durkan2019neural} constitute a
class of efficient and flexible density estimators for modeling
complex probability distributions. This makes it particularly
effective for NPE. In this work, we use normalizing flows to
construct posterior estimators, with the detailed architecture
shown in Fig.~\ref{fig:model}.

\begin{center}
\refstepcounter{figure}
\includegraphics[width=0.8\textwidth]{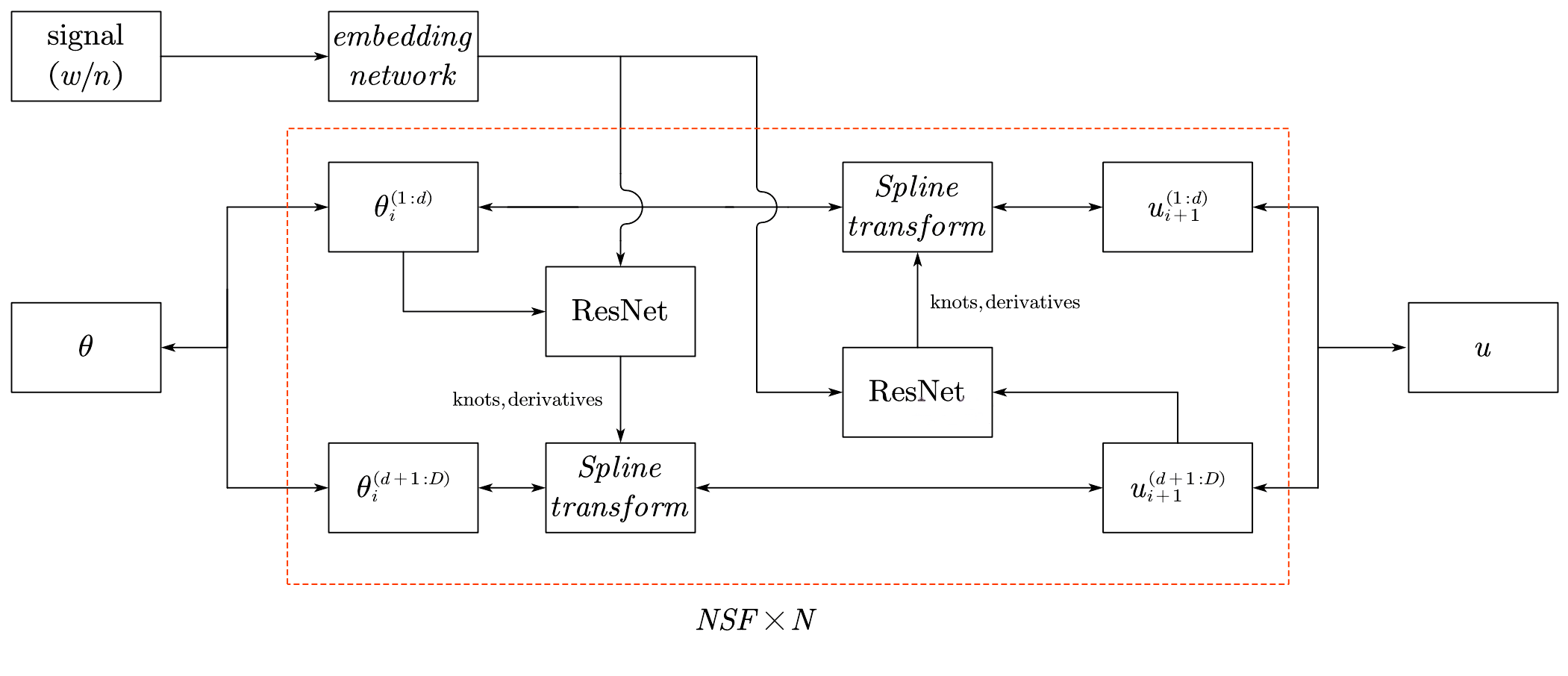}

\textbf{Figure \thefigure.}
The architecture of our machine learning framework.
\label{fig:model}
\end{center}

To exploit underlying properties such as equivariances in the
parameter space, group equivariant neural posterior estimation
(GNPE)~\cite{dax2021group} was proposed. It allows us to
incorporate the joint symmetries of data and parameters into
neural network architecture, thereby simplifying the learning task
for the neural network.

In GW parameter estimation, a shift in the coalescence time $t_c
\rightarrow t_c+\Delta t$ corresponds to a phase rotation in the
frequency-domain strain data,
\begin{align}
d(f)\rightarrow d(f)e^{-2\pi i f\Delta t}.
\end{align}
Under this transformation, the posterior distribution is equivariant,
\begin{align}
p(t_c|d)=p(t_c+\Delta t\,|\,d\,e^{-2\pi i f\Delta t}).
\end{align}
%Here, we take the pose proxy $\hat t_I = t_I + \epsilon_I$, where
%$\epsilon \sim Unif(-1ms,1ms)$. GNPE utilizes such equivariances
%to transform the data into a standardized representation, reducing
%the complexity of the learning task and improving the efficiency
%of posterior inference.

\section{Results}
\label{sec:HBA}

To apply GNPE to GW inference, two separate neural networks are
trained in the \texttt{dingo}~\cite{dax2021real,dax2021group}
framework： an initialization network that estimates the pose for Gibbs sampling and a main posterior network that learns the conditional posterior distribution.
%The first is an initialization network that models the proxy-variable distribution. Its purpose is to provide an initial estimate of the pose (the coalescence times), which serves as the starting point for the Gibbs sampler. %Since this network only needs to infer two or there parameters and does not require highly accurate predictions, it can be significantly smaller than the full posterior model.
%The second network is the main GNPE posterior model, which learns the conditional posterior distribution Eq.~(\ref{eq:gibbs}). %In this formulation, the detector data are implicitly transformed according to the inferred proxy variables, such that the exact time-translation symmetry of the gravitational-wave signal is enforced during inference.

\begin{center}
\refstepcounter{figure}
\includegraphics[width=0.8\textwidth]{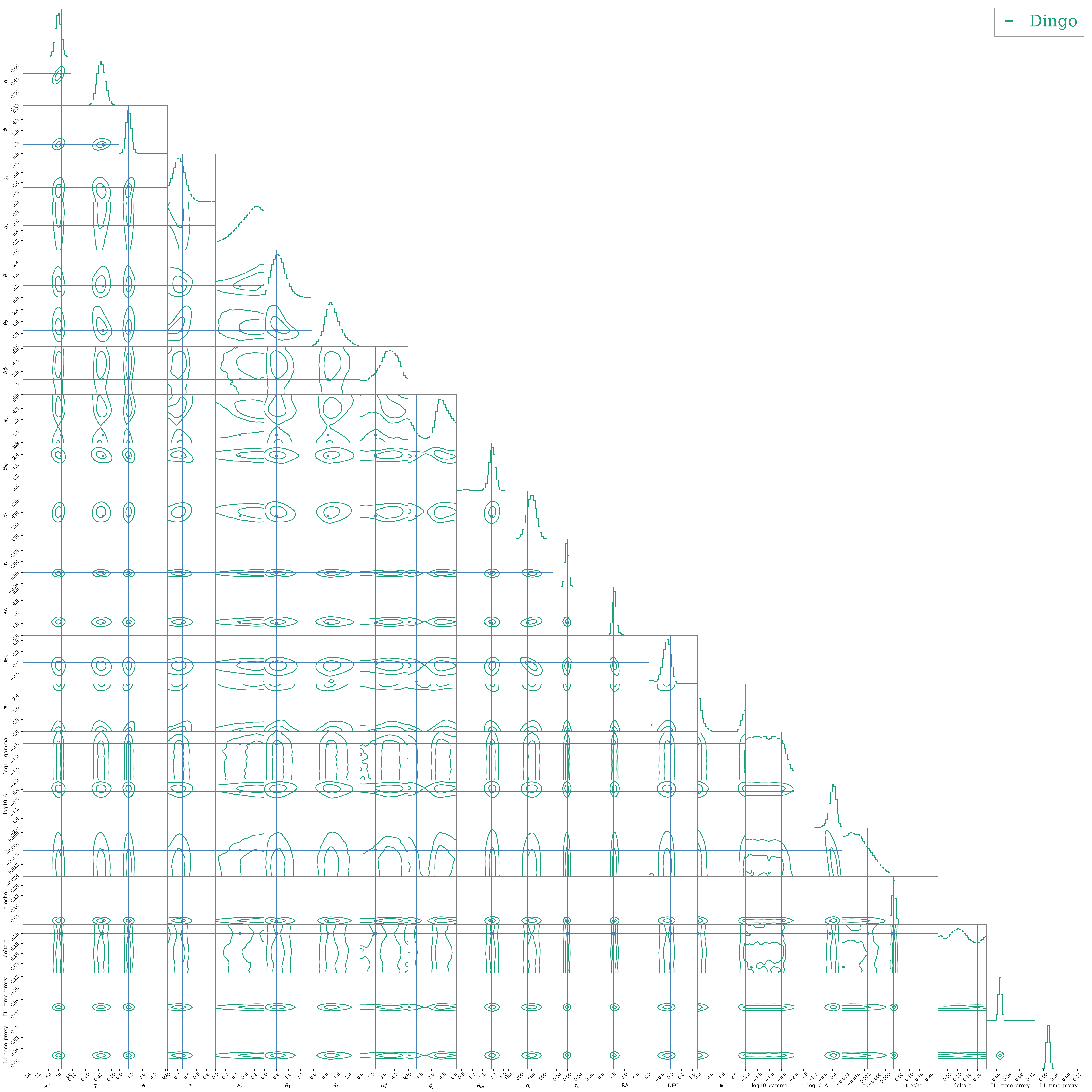}

\textbf{Figure \thefigure.}
Posterior distributions for a single injected gravitational-wave signal. The blue lines indicate the injected parameter values.
\label{fig:corner}
\end{center}

The initialization network and the main GNPE network can be
trained simultaneously, with each network running on a separate
RTX 5090 GPU and the full training procedure typically requiring
about 9--10 days. Once trained, the model could be directly
applied to GW echo searches. The inference for a single event
required less than 20 minutes, whereas traditional Bayesian
inference methods typically require several hours, resulting in a
significant improvement in computational efficiency.

To validate the inference performance of the model, we injecte a relatively prominent echo signal ($log_{10}A = -0.5$) and perform inference on it, as shown in Fig.~\ref{fig:corner}. From the posterior distribution, it can be seen that our model can detect the echo well.

Then, we perform
posterior inference on 1000 simulated data sets and construct a
P–P plot, as showed in Fig.~\ref{fig:pp_plot}. For each
parameter we compute the percentile score of the true injected
value within the corresponding marginalized posterior
distribution, and construct the cumulative distribution function
(CDF) of these percentiles. For an unbiased posterior inference
method, the percentiles should follow a uniform distribution,
corresponding to the diagonal line in the P--P plot. The legend
shows the p-values of the individual parameters, with a combined
p-value of 0.154. This shows that the GNPE is performing properly
on simulated data

\begin{center}
\refstepcounter{figure}
\includegraphics[width=0.8\textwidth]{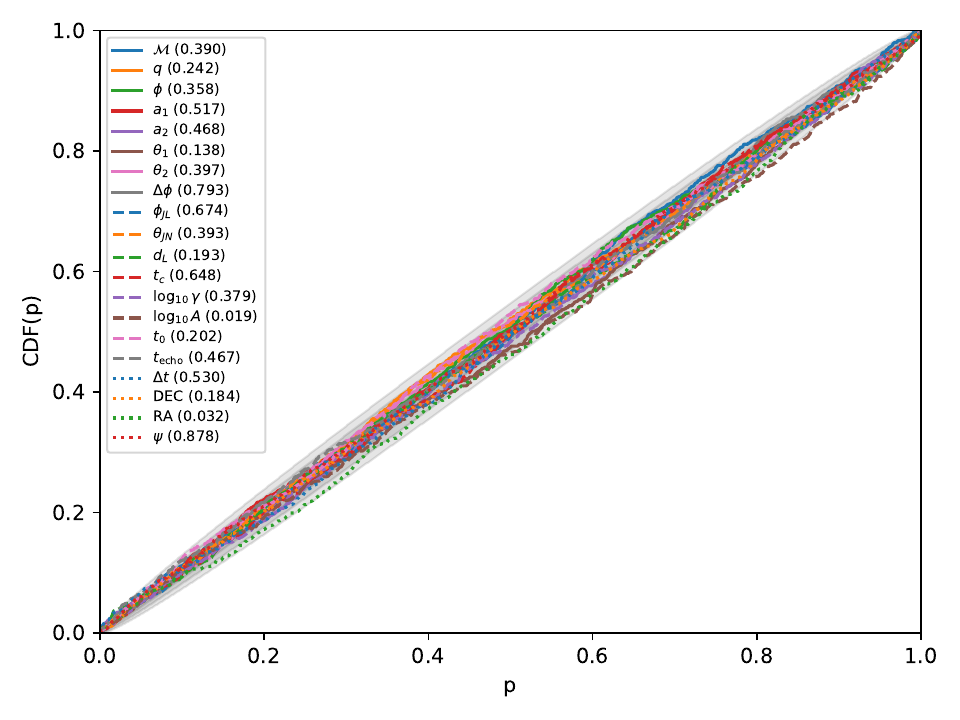}

\textbf{Figure \thefigure.} Probability-Probability plot for a set
of 1000 posterior evaluations. The legend shows the p-values of
the individual parameters, with a combined p-value of 0.154.
\label{fig:pp_plot}
\end{center}

%Figure~\ref{fig:corner} shows the posterior inference result for a randomly injected signal. The blue vertical lines indicate the injected parameter values. It can be seen that GNPE is able to accurately recover all parameters within the posterior distributions.

We next perform an echo search for the O4a GW events. After
posterior inference, we compare the IMRPhenomXPHM+Echo(IMRE) and
the IMRPhenomXPHM(IMR) waveform hypothesis by computing Bayes
factors. Importance sampling was used when computing Bayes factors. For a given hypothesis $\mathcal{H}_i$, the Bayes factor
against the noise-only hypothesis $\mathcal{H}_0$ is defined as
\begin{align}
B_i=\frac{P(d\mid\mathcal H_i)}{P(d\mid\mathcal H_0)},
\end{align}
where $P(d\mid\mathcal H_i)$ is the Bayesian evidence for
hypothesis $\mathcal H_i$, and $P(d\mid\mathcal H_0)$ is the
evidence for the noise-only hypothesis. For the importance sampling,  we used 50,000 samples. From our results, the sampling efficiency is roughly around $10\%$, and the uncertainty in the evidence is approximately 0.01. The results are summarized
in Table~\ref{tab:result}, where
\begin{align}
\log_{10}B^{\mathrm{IMRE}}_{\mathrm{IMR}}
=
\log_{10}\frac{B^{\mathrm{IMRE}}}{B^{\mathrm{IMR}}}.
\end{align}

\begin{table}[H]
\centering
\caption{Bayes factors for O4a events.}
\begin{tabular}{lr@{\hspace{1cm}}lr}
\toprule
Event & $\log_{10} B^{\mathrm{IMRE}}_{\mathrm{IMR}}$ & Event & $\log_{10} B^{\mathrm{IMRE}}_{\mathrm{IMR}}$ \\
\midrule
GW230606\_004305 & -1.241 & GW230831\_015414 & -1.411 \\

GW230624\_113103 & -1.274 & GW230904\_051013 & 0.174 \\

GW230630\_125806 & -0.792 & GW230930\_110730 & -2.078 \\

GW230704\_021211 & -2.770 & GW231005\_091549 & -1.424 \\

GW230707\_124047 & -1.646 & GW231026\_130704 & -1.213 \\

GW230708\_053705 & -1.628 & GW231113\_122623 & -2.907 \\

GW230803\_033412 & -3.235 & GW231113\_200417 & 0.864 \\

GW230805\_034249 & -1.610 & GW231127\_165300 & -2.564 \\

GW230806\_204041 & -0.870 & GW231129\_081745 & -3.460 \\

GW230825\_041334 & -2.308 & GW231223\_032836 & -3.470 \\
\bottomrule
\end{tabular}

\label{tab:result}
\end{table}

For the IMR waveform inference, we utilize the PyCBC
framework~\cite{allen2012findchirp,Allen:2004gu,Nitz:2017svb,DalCanton:2014hxh,Usman:2015kfa}.
During the inference process, the data segment, PSD estimate, frequency range, sampling rate, waveform settings (modes, reference frequency), and priors are kept exactly the same as those used in
the IMRE waveform analysis. The values of
$\log_{10}B^{\mathrm{IMRE}}_{\mathrm{IMR}}$ around or below zero
indicate no significant evidence for GW echo. 

For the two events with positive Bayes factors, the posterior distributions of their echo parameters are close to the prior distributions, as shown in Figure~\ref{fig:echo_posteriors}, indicating that no significant echo signal exists. Therefore, we find no significant evidence for GW echo in the analyzed O4a events.

\begin{center}
\refstepcounter{figure}

\begin{minipage}{0.48\textwidth}
    \centering
    \includegraphics[width=\textwidth]{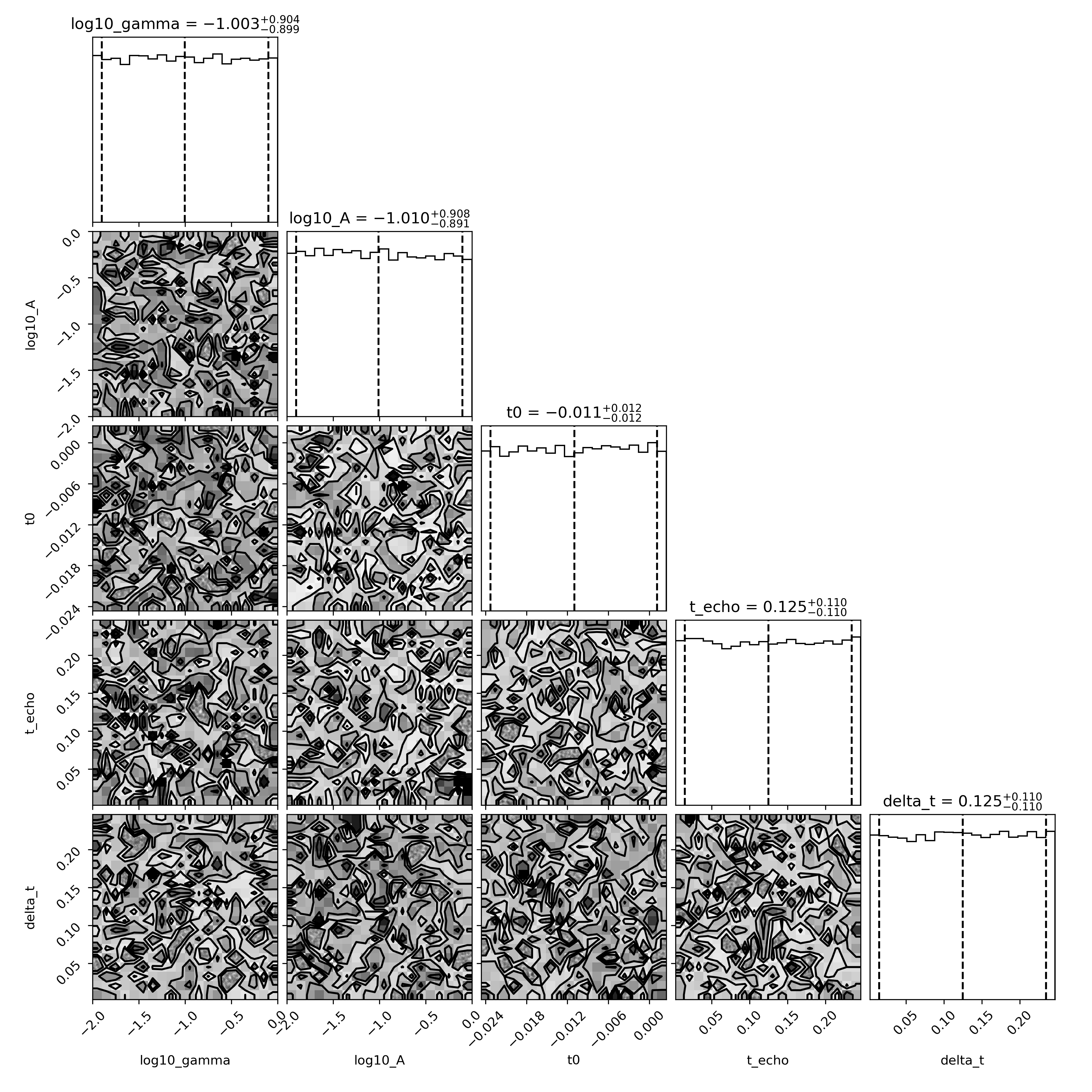}
\end{minipage}
\hfill
\begin{minipage}{0.48\textwidth}
    \centering
    \includegraphics[width=\textwidth]{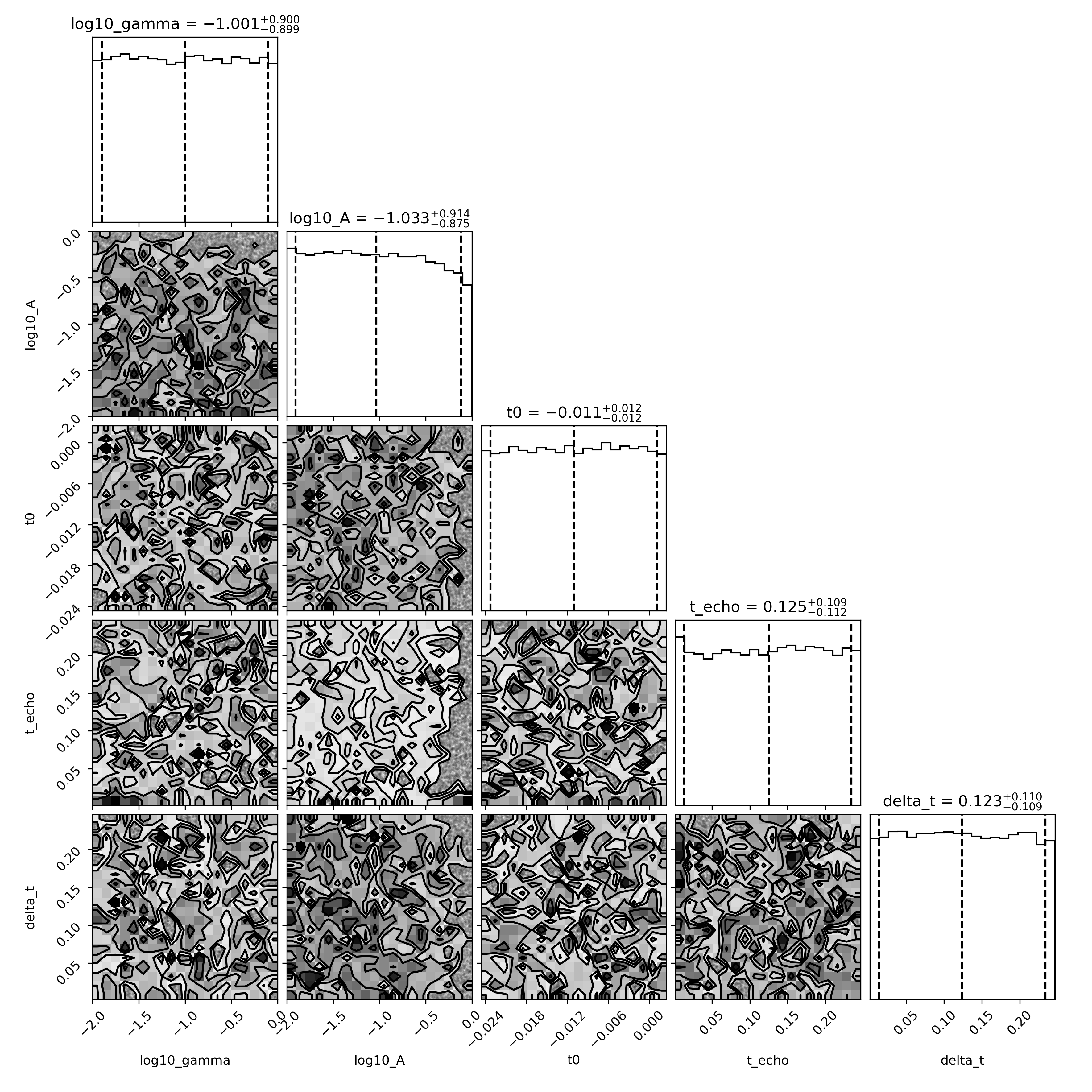}
\end{minipage}

\textbf{Figure \thefigure.} Posterior distributions of the echo
parameters for GW230904\_051013 (left) and GW231113\_200417 (right).
\label{fig:echo_posteriors}

\end{center}

\FloatBarrier

\section{Conclusion}

%In this Letter, we employ GNPE to construct a model for searching gravitational-wave echoes. After the training procedure was completed, we performed injection tests and analyzed O4a events. The results demonstrate that the method is capable of accurately inferring all parameters. For the O4a events, we find that , indicating that no evidence for the presence of gravitational-wave echo signals is found.

In this work, we employ GNPE based on normalizing flows to
construct a model for searching for GW echoes. The model can
perform Bayesian inference for a single GW event in less than 20
minutes, compared with several hours required by traditional
Bayesian inference methods. This substantially improves the
efficiency of GW echo searches while significantly reducing the
computational cost.

%by incorporating the time-translation symmetry of
%gravitational-wave signals into the network architecture.

%The framework consists of an initialization network and a main
%GNPE network. The initialization network provides an initial
%posterior estimate, while the main GNPE network iteratively
%refines the posterior distribution. Both networks were trained
%using a dataset of $5\times 10^6$ waveforms generated with the
%IMRPhenomXPHM+Echo waveform model, to enable efficient Bayesian
%inference of GW echo signals. The two networks were trained
%simultaneously on two RTX 5090 GPUs, with the complete training
%procedure requiring 9--10 days. After training,

After validating the inference performance of the model on
injected signals, we performed a search for
the echo in O4a events. The results are summarized in
Table~\ref{tab:result}. For the majority of the analyzed events,
the values of $\log_{10}B^{\mathrm{IMRE}}_{\mathrm{IMR}}$ are
below zero, indicating that the data favor the IMRPhenomXPHM
waveform model without echo over the IMRPhenomXPHM+Echo model.
Although two events, GW230904\_051013
($\log_{10}B^{\mathrm{IMRE}}_{\mathrm{IMR}}=0.174$) and
GW231113\_200417
($\log_{10}B^{\mathrm{IMRE}}_{\mathrm{IMR}}=0.864$), yield
positive Bayes factors, the support for the echo hypothesis is
weak and is not statistically significant. Overall, we find no
significant evidence for the GW echo in the analyzed O4a events.

%\section{Supplementary Materials}

\section*{Acknowledgments}

This work is supported by National Key Research and Development
Program of China, No. 2021YFC2203004, and the Fundamental Research
Funds for the Central Universities.

\bibliography{REF}

\end{document}